\documentclass[
twocolumn, preprintnumbers, amsmath,amssymb,nofootinbib]{revtex4}
\usepackage{graphicx}
\usepackage{dcolumn}
\usepackage{bm}
\usepackage{amsmath,amscd}

\newcommand\ee{\end{equation}}
\newcommand\be{\begin{equation}}
\newcommand\eea{\end{eqnarray}}
\newcommand\bea{\begin{eqnarray}}
\newcommand{\Dsl}{\slash \!\!\!\!D}

\begin{document}


\title{Scalar-tensor theories, trace anomalies and the QCD-frame}

\author{Francesco Nitti$^1$ and Federico Piazza$^{1,2}$}

\affiliation{%
$^1$Laboratoire APC, Universit\'e Paris 7, 75205 Paris, France\\
$^2$Paris Center for Cosmological Physics, Universit\'e Paris 7, 75205 Paris, France
}%


\begin{abstract}
We consider the quantum effects of matter fields in scalar-tensor theories and clarify the role of trace anomaly when switching between conformally related `frames'. We exploit the property that the couplings between the scalar and the gauge fields are not frame-invariant in order to define a `QCD-frame', where the scalar is not coupled to the gluons. We show that this frame is a natural generalization of the `Jordan frame' in the case of non-metric theories and that it is particularly convenient for gravitational phenomenology: test bodies have trajectories that are as close as possible to  geodesics  with respect to such a metric and equivalence principle violations are directly proportional to the scalar coupling parameters written in this frame.  We show how RG flow and decoupling work in metric and non-metric theories. RG-running commutes with the operation of switching between frames at different scales.  When only matter loops are considered, our analysis confirms that  metricity is stable under radiative corrections and shows that approximate metricity is natural in a technical sense.  
\end{abstract}

\maketitle

\section{Introduction} \label{sec:1}
\noindent

The discovery of the acceleration of the Universe has triggered a renewed interest in scalar-tensor theories of gravity. Many models recently proposed (\cite{DGP,justin,f(r),DHK,galileon} just to mention a few), involve, at least in some limit, a light scalar degree of freedom universally coupled to matter.  The universality of the couplings is what defines a \emph{metric theory}~\cite{BD}, i.e. a theory in which matter couples minimally to a unique metric tensor ${\tilde g}_{\mu \nu}$. Schematically, 
\begin{equation} \label{action}
S = {M_*^2\over 2}\int  \sqrt{\tilde g} \  \Omega^{-2}(\phi) \tilde R +  S_\phi [\tilde g, \phi]\, + S_m[\tilde g, \psi_i] \, , 
\end{equation} 
where $\tilde R$ is the Ricci scalar of $\tilde{g}_{\mu\nu}$,  $S_\phi$ contains kinetic and self-interaction terms for $\phi$\footnote{Most of the analysis of the present paper applies regardless of the type of scalar self-interaction. However, when we report experimental bounds on the scalar matter couplings and study the equivalence principle phenomenology in Sec.~\ref{sec:3} we simply assume a sufficiently light minimally coupled field and no screening mechanism at work}, and $S_m$ is the action for matter fields $\psi_i$. These  couple to the scalar $\phi$  only through the `Jordan' (J-) metric ${\tilde g}_{\mu \nu}$. As a result, test particles, independently of their compositions, follow trajectories that are geodesics of ${\tilde g}_{\mu \nu}$. On the other hand, 
the effective Newton constant depends on $\phi$, thus  the universality of free fall is violated by~\eqref{action} only in its `strong' form, i.e. among bodies of different -- and non negligible -- gravitational self-energies. The equivalence principle (EP)~\cite{D,W}  is not otherwise violated by metric theories. 

String theory, with its dynamical couplings and omnipresent moduli fields, offers another main motivation for looking at scalar-tensor theories~\cite{TV}.
The string \emph{dilaton} can benefit from a `decoupling mechanism' at a symmetric point in field space~\cite{DP}, or at infinity~\cite{DPV} that suppresses EP violations below the present experimental limits (but very close, in the case of~\cite{DPV}). However, metric dilaton couplings like in~\eqref{action} look unnatural in string theory because they are not present at tree-level to begin with.   
Another possible source of uneasiness with theories like~\eqref{action} is that metricity could be spoiled by loop corrections and/or compositeness. To which extent and in which cases this is true has been the matter of some debate. One motivation of the present paper is to further clarifying this issue.

It is often convenient to use a metric $g_{\mu\nu}$ that is conformally related to ${\tilde g}_{\mu \nu}$, by $\tilde{g}_{\mu\nu} = \Omega^2(\phi) g_{\mu\nu}$. In the resulting  `Einstein' (E-) frame  the gravitational action takes the Einstein Hilbert form:
\begin{equation} \label{emetric}
S = \frac{M_*^2}{2} \int \sqrt{g} \, R\  + S^{(E)}_\phi[g,\phi] + S^{(E)}_m[g, \phi, \psi_i]\, .
\end{equation} 
Note that we can always define an E-frame but only metric theories allow a J-frame. 
In the E-frame, the propagators of the graviton and of the scalar field decouple and the contributions of different bodies to the total gravitational field are found straightforwardly.  However, the J-frame looks more transparent in several other respects. 

First of all, the J-frame is the one directly measured by clocks, rods etc\dots, in the approximation that gravity is negligible inside such measuring devices. We devote App.~\ref{app_a} to clarify this point. 
Beside -- and related with -- the above consideration, there is a number of properties that, from a J-frame viewpoint,  look almost self-evident: 
\begin{enumerate}
\item As mentioned, test bodies follow geodesics; therefore, (weak) EP is not violated. 
\item Couplings are not varying (at least they are not varying with $\phi$!). 
\item Metricity is preserved no matter how we assemble objects together, as far as the gravitational self-energy of the composite remains negligible: metricity is stable against compositeness. 
\item Metricity is stable under radiative corrections \emph{as long as} only matter loops are considered (and UV-regulators are used which are $\phi$-independent -- it is natural to do so in the J-frame). Gravitons and scalar loops are expected to produce Planck suppressed -- but not necessarily unobservable -- deviations from metricity. 
\end{enumerate}

The last two points can easily be justified by considering the quantum path integral in the Jordan frame, which we will do shortly below.
Thus, in metric theories, matter  quantum corrections and  compositeness cannot lead to violations of the weak equivalence principle: as long as one can safely consider the metric and dilaton as a fixed background, all bodies follow trajectories which are geodesic of the given J-frame metric. This conclusion is straightforward in the J-frame, but it looks far from trivial  if one starts with the same theory written in the Einstein frame, where it  will seem as arising from several cancellations between different contributions. This was confirmed by  explicit E-frame calculations~\cite{F,HN,A}, and we will show its validity in Sec.~\ref{sec:4} by calculating the renormalization group flow of the scalar couplings.

On the other hand, including the dynamics of $\phi$ and $g$ is generically expected to break metricity at the quantum level. 
This suggests that models recently presented as metric theories could in fact be non-metric, but only slightly so, metricity being violated by Planck suppressed operators~\cite{A}.  For such `quasi-metric' theories, it is useful to define a `quasi J-frame', in virtue of all the good qualities above mentioned.

The main purpose of this note is to propose a definition for such a frame. We do that in Sec.~\ref{sec:3}.
We will show that even in a generic non-metric theory, it is possible to define a  frame  in which, at least for massive test bodies, the trajectories are ``as close as possible'' to geodesics of an appropriate metric. We dubbed this metric `QCD-frame' because it is the one in which the scalar field does not couple to the gluons. We show that, in this frame, it is straightforward to parametrize the effects of the dilaton in experiments that measure the {\em difference in  acceleration} between test bodies falling in a common gravitational field generated by a very massive third object. 

At the basis of our `QCD-frame' idea is the possibility of setting the dilaton-gluon couplings to zero by an appropriate conformal transformation. Such a possibility is guaranteed by trace anomaly, that makes the dilaton couplings to gauge fields frame-dependent. As this simple observation has been seemingly overlooked by some of the existing literature on scalar-tensor theories, we devote Sec.~\ref{sec:2} to clarify the role of trace anomaly in switching between conformal frames. 

Finally, in Sec.~\ref{sec:4} we show that not only is metricity stable under radiative correction, but that also {\em approximate} metricity shares the same property, in the sense that a small deviation from metricity does not become large as  a result of quantum corrections. In other words, the smallness of weak EP violation effects, although non-generic a priori from a fundamental theory point of view, is technically natural.

In all that follows we will be interested only in the {\em linear} dilaton couplings to matter. Therefore  we will consider conformal factors $\Omega(\phi)$ of the form
\be\label{linear}
\Omega(\phi(x)) = \Omega(\phi_0 + \delta\phi (x))  =   1 + \alpha \varphi(x)
\ee
where $\alpha=\Omega'/\Omega$ is a constant and $\varphi(x)$ can be considered as a small deviation from an otherwise constant dilaton background $\phi = \phi_0$ which we can set to $\phi_0=1$ by a redefinition of the Planck mass in the Einstein frame. 

Before we procede further, we provide a simple path integral argument to show that metricity is radiatively stable under loops in the matter sector. 
If we treat the metric and dilaton as external sources, and integrate over some of the matter fields $\psi_i$, the resulting effective action for the metric, dilaton and remaining matter fields $\Psi_i$ will read 
\be\label{ZJ}
e^{i S_{eff}[g,\phi,\Psi_i] }= \int {\cal D}[\psi_i] \,e^{iS_m[\tilde g, \psi_i, \Psi_i]}\,  
\ee
Since the right hand side is a functional of $\tilde{g}$ alone, the  effective action will depend only on the Jordan  frame metric, and not on the dilaton independently, i.e. 
$$
S_{eff}[g,\phi,\Psi_i] \equiv S_{eff}[\tilde{g},\Psi_i].
$$
Similarly, the quantum generating functional for matter Green's functions, obtained by adding external sources $J_i$ for the matter fields, will only depend on $\tilde{g}$  and not on $g$ and $\phi$ independently, 
$Z[g,\phi,J_i] \equiv Z[\tilde{g},J_i]$.    
Thus, all loop corrections will involve only universal coupling to the J-frame metric and will not violate the weak equivalence principle. This includes the renormalization procedure, as long as in the renormalization scheme all scales and momenta are kept fixed in the J-frame, i.e. they do not depend explicitly on $\phi$. 

The same holds for the Wilsonian  effective action for any low-energy degree of freedom and composite states. If we think of $S[\tilde{g},\psi_i]$ as the effective action for the fundamental degrees of freedom at a scale $\Lambda$, the Wilsonian effective action for the light d.o.f. $\Psi_i$ at a scale $\mu < \Lambda$ is obtained by  integrating out the heavy d.o.f.  above the scale $\mu$: schematically, 
\be
e^{iS_{eff}^{(\mu)}[\tilde{g},\Psi_i] } =\int_{\mu^2 < p^2 < \Lambda^2}  {\cal D}[\psi_i(p)]\, e^{iS_m^{(\Lambda)}[\tilde g, \psi_i,\Psi_i]}\, . 
\ee
As the functional integral over the heavy d.o.f. is a function of $\tilde{g}$ alone, one can see immediately that, as long as the IR and UV cutoff are defined in a $\phi$-independent way, metricity is preserved, the  low energy fields in the effective action couple to the J-frame metric only, and explicit couplings  to the dilaton are absent at any scale.

 \section{Metric theories in the E-frame}
\label{sec:2}

Before considering general non-metric theories, it is worth clarifying the role of trace anomaly in going from one frame to a conformally related one.  Starting from a metric theory  in the J-frame, this  will allow us to derive the linear dilaton couplings to matter in the Einstein frame. 

Consider a metric theory, whose effective action at a scale $\Lambda$ contains fermions and gauge fields (abelian and non-abelian). For example, one can think of the $SU(3) \times U(1)$ gauge theory with fermionic matter which decribes  the Standard model at scales less than $\sim 100$ GeV. The matter action  in the J-frame is 
\bea 
S_m = \int \sqrt{\tilde{g}}\!\!\!\! && \left[\sum_i\bar{\psi}_i \left (i \Dsl - m_i \right) \psi_i \, +\right. \nonumber \\  && \left. -{1\over 4e^2} F_{\mu\nu}F^{\mu\nu} \, - \, {1\over 4 g_3^2} F^a_{\mu\nu}F^{a\, \mu\nu}  \right]
\label{Jframe}\eea
The indices are contracted with $\tilde g^{\mu\nu}$, and by assumption the strong and electromagnetic couplings, $g_3$ and $e$, as well as the fermion masses, are $\phi$-independent. 
Based on  classical conformal transformations, one could argue that linear  scalar couplings to the gauge fields never arise in {\em any} frame, since the gauge field action is invariant under Weyl rescalings of the metric. By the same argument, one could think that explicit dilaton couplings to the gauge fields  are intrinsically non-metric in four dimensions: if present in one frame, they will evaluate the same in all conformally related frames. 

However, this argument is invalid: conformal invariance is  broken  at the quantum level, due to the presence of the trace anomaly. As noticed in \cite{brax}, this leads to the emergence of linear dilaton couplings to gauge fields in going from one frame to another. As one can see from the explicit calculation in  \cite{brax}, the anomaly arises as a Jacobian  in the path integral measure over the fermions: If we rewrite the action in equation (\ref{Jframe}) explicitly in terms of  $g_{\mu\nu}$ and $\Omega$, the fermionic action reads:
\be
S_\psi = \int \sqrt{g}\sum_i \Omega^3\bar{\psi}_i \left (i \Dsl - \Omega m_i \right) \psi_i, 
\ee 
and to reabsorb the scale factor $\Omega^3$ in the  kinetic term  (so that it is canonical in the Einstein frame) one has to make  the change of variables  $\psi_i' = \Omega^{3/2}\psi_i$. This produces the anomalous Jacobian. 

Here, rather than going through the explicit calculation, we illustrate this  by simply  highlighting the general relation between E-frame gauge  couplings and the trace anomaly, and use the well-known field theory result for the latter.  Let us write the path integral as a function of the background fields $g_{\mu\nu}$ and $\varphi$. The generating functional depends only on the J-frame metric $\tilde{g}_{\mu \nu} = \Omega^2 g_{\mu \nu}$,  
\be
Z[g,\varphi] \equiv Z[\Omega^2(\varphi) g] = \int {\cal D}[\psi_i] {\cal D}[A^a_\mu]  e^{iS_m[\Omega^2 g_{\mu \nu}, \psi_i, A_\mu^a]} \, .
\ee
Differentiating with respect to $\varphi$ and evaluating at $\varphi=0$ we obtain an insertion of the trace of the stress tensor:
\bea
{\delta Z[g,\varphi] \over \delta \varphi(x)}\Big|_{\varphi=0} \!\! = && \!\!\!\!\! i \int {\cal D}[\psi_i] {\cal D}[A_\mu]   \, {\delta S_m \over  \delta \tilde g_{\mu\nu}} g_{\mu\nu}
{ d\Omega^2 \over d\varphi} e^{iS_m} \nonumber \\ 
= && \!\!\!\!\! i \alpha \int {\cal D}[\psi_i] {\cal D}[A_\mu]    \sqrt{\tilde g}\, T_\mu^\mu (x) e^{iS_m} \label{dZj}
\eea 
where we have used eq. (\ref{linear}) and the definition of the (J-frame-) stress tensor, $T^{\mu\nu}= (2/\sqrt{\tilde g}) (\delta S_m / \delta \tilde g_{\mu\nu})$. We can also perform the same calculation in the Einstein frame, in which at the linear level in $\varphi$ the action will take the general form: 
\be\label{Eframeaction}
S^{(E)}_m[g,\varphi; \psi_i,A_\mu] = S_m[g,\psi_i,A_\mu] + \int \sqrt{g} \varphi(x) {\cal A}(x)
\ee
where $S_m[g,\psi_i,A_\mu]$ is the same action as in the J-frame\footnote{To be precise, as mentioned before, the canonically normalized fermions in the E-frame are related to those in the J-frame by $\psi^{(E)}_i = \Omega^{3/2} \psi_i$. Below we omit the label $(E)$  and rename the Einstein fermions $\psi_i$. This relabeling is immaterial when writing the path integral, as long as one keeps track of the appropriate Jacobian, which precisely gives rise to the extra anomaluos contribution in eq. (\ref{Eframeaction}).}, but with $\tilde g \to  g$. 
The matter functional integral in terms of the Einstein frame action reads:
\be
Z[g,\varphi] =\int {\cal D}[\psi_i] {\cal D}[A^a_\mu]  e^{iS_m^{(E)}[g_{\mu \nu}, \varphi,\psi_i, A_\mu^a]} \, .
\ee
Once again, differentiating with respect to $\varphi$ we obtain: 
\be\label{dZe}
{\delta Z[g,\varphi] \over \delta \varphi(x)}\Big|_{\varphi=0} =  i\int {\cal D}[\psi_i] {\cal D}[A_\mu]  e^{iS_m[g,\psi_i,A_\mu]} \sqrt{g} {\cal A}(x). 
\ee
On the other hand the frame change is nothing but a field redefinition in the path integral variables, so the two results (\ref{dZj}) and (\ref{dZe}) must agree. The same must be true if we insert any combination of the fields  in the path integral. This imposes the relation 
\be 
{\cal A}(x) = \alpha T_\mu^\mu (x) \, .
\ee
In other words, the additional term needed in the Einstein frame action (\ref{Eframeaction})  must coincide with the quantum field theory  trace anomaly. This includes a classical contribution from the fermion mass terms, plus a quantum contribution containing  the beta-functions of the theory and the fields' anomalous dimensions. The full linear dilaton coupling in the Einstein frame thus reads   
\bea
S^{(E)}_m = S_m + && \!\!\!\!\!\alpha \int \sqrt{g} \,\, \varphi  \left[\sum_i m_i \left( 1 + \gamma_{m_i} \right) \bar{\psi}_i\psi_i \, +\right. \nonumber \\  && \!\!\!\!\!\!\!\! \left. +\,  {\beta(e)\over 2e^3} F_{\mu\nu}F^{\mu\nu} \, + \, {\beta_3(g_3)\over 2g_3^3} F^a_{\mu\nu}F^{a\, \mu\nu}  \right], \label{fullaction}
\eea
where $\beta_3(g_3)$ and $\beta(e)$ are the QCD and QED $\beta$-functions and 
$\gamma_i(g_3,e)$ are the fermions anomalous dimensions.  
As we said earlier, we  will be considering this action in a Wilsonian sense. In this case, if we perfom the frame change at a certain scale $\Lambda$, we should consider the $\beta$-functions contributions of the relevant degrees of freedom at that scale. This may seem counter-intuitive, but as we will explicitly show in Section~\ref{sec:4} it  is a  consequence of the underlying  metricity of the theory and  of the decoupling of heavy fields at scales lower than their masses. 

A few comments are in order. 
Firstly, the theory in the Einstein frame looks fine-tuned, since all the dilaton couplings are related in a very specific way and depend on the single parameter $\alpha$. However, we know from the J-frame analysis of Sec.~\ref{sec:1} that the underling metricity prevents this tuning to be destroyed  by quantum corrections: thus the form of the action (\ref{fullaction}) is radiatively stable.  

Secondly, the presence of a scalar-photon coupling $\varphi F_{\mu \nu}F^{\mu \nu}$ does not imply \emph{per se}, EP-violations nor potential coupling variations. For instance, if all the other terms are `tuned' as in~\eqref{fullaction} the theory is perfectly metric and the four properties mentioned in the introduction apply. This should be taken into account when making connection with phenomenology. 
Experimental limits on EP violations are normally used (e.g.~\cite{DP,DPV,KW,DZ,dent,DD}) to bound the E-frame QED dilaton coupling $\tilde d_e$, defined  by\footnote{Whenever convenient, we use the notation of~\cite{DD}. We make an exception by calling $\tilde d_e$ what in~\cite{DD} is called $d_e$. Instead, for us $d_e$ is defined  as $d_e = - e \, \tilde d_e/(2 \beta)$ (see next section).} 
 \begin{equation}  \label{defi}
{\cal L}^{(E)}_{QED} = - \frac{1}{4 e^2}\left(1 - \tilde d_e  \, \varphi \right) F_{\mu \nu} F^{\mu \nu}\, . \\
\end{equation}
However, as we show below, the value of $\tilde d_e$ for a metric theory can exceed the experimental bounds that are normally quoted when the effects of QED trace anomaly are neglected. 

What is the largest value of $\tilde d_e$ compatible with a metric theory? From~\eqref{fullaction} $\tilde d_e = 2 \alpha \beta(e)/e$. At the nuclei scales $\sim 1$ GeV relevant for EP experiments we can safely calculate $\beta(e)$ by considering electron loops, giving 
\begin{equation} \label{d_e}
\tilde d_e  \simeq \frac{\alpha \, e^2}{6 \pi^2} .
\end{equation}
This is likely to be a lower bound because the positive contribution of the muon to the beta function  has been neglected.
Now, the current upper bound on composition-independent effects gives $\alpha^2 < 10^{-5}$~\cite{bertotti}. Therefore, according to~\eqref{d_e}, $\tilde d_e$ can be as large as $\tilde d_e \sim 5 \times 10^{-6}$.
On the other hand, by using the combined results of the E\"otWash collaboration and of Lunar Laser Ranging experiments,  the upper bound $|\tilde d_e \alpha | \lesssim 4 \times 10^{-9}$ is found~\cite{DD}, which translates into $|\tilde d_e| \lesssim 1.2 \times 10^{-6}$ for the maximum allowed value of $\alpha$. We are going to show that what EP experiments actually constrain is, more properly, the `QCD-frame' dilaton couplings $c_e$ that we define in Section~\ref{sec:4}.

\section{Non-metric theories and the QCD-frame} \label{sec:3}

Although metric theories display many nice properties, there is no reason a priori  that the low energy effective action be of this kind. As mentioned, string theory does not seem to have any particular preference for metricity. But also standard model portals into light sectors (e.g.~\cite{Carroll,PP}) will generally violate the universality of the couplings. Thus, let us now turn to general non-metric theories, in which  there is no frame where  all matter-dilaton couplings vanish.
In the last section we noted that  gauge dilaton couplings in the effective action are frame-dependent. Therefore,  we can exploit the trace anomaly trick and reabsorb to zero {\em one}  coupling of our choice  in some appropriate frame.
As most of the mass of a proton or a neutron comes from the gluonic contribution $\langle TrF^2\rangle$, in order to describe the interaction of ordinary matter it is natural to introduce  {\em QCD-frame}  where the scalar does not couple directly to gluons at the level of the effective action. 

In writing the QCD-frame action we follow closely the remarkable analysis of Damour and Donoghue~\cite{DD}. There, the E-frame dilaton couplings and the relative charges are discussed and calculated. The authors argue that the effective action relevant for EP experiments is that at the nucleon energy scale,  $\Lambda_{QCD}\sim 1$ GeV, and includes only the QED and QCD gauge fields, the two lightest quarks and the electron fields. Heavier fields are meant to be already integrated out, while the contribution of the strange quark is argued to be negligible. 

If we start with the above field content and generic dilaton couplings, in the Einstein frame  the Lagrangian will read, in the notation of \cite{DD}:
\begin{align} \nonumber
{\cal L}^{(E)}_ m = &- \frac{1}{4 g_3^2}\,\left(1 + d_g\,{2 \beta_3\over g_3} \varphi\right) F^a_{\mu \nu} F^{a\, \mu \nu} \\ & - \frac{1}{4 e^2} \left(1 + d_e \, {2 \beta \over e}\varphi\right) F_{\mu \nu} F^{\mu \nu} \label{efull} \\ 
& + \!\! \sum_{i=u,d,e}\left[i{\bar \psi}_i \, {\slash \!\!\!\!D} \, \psi_i - m_i(1+ d_{m_i}\varphi + d_g \gamma_{m_i} \varphi) {\bar \psi}_i \psi_i\right] . \nonumber
\end{align}
where the $d_i$ parameters are  chosen in such a way as to multiply renormalization group invariant operators\footnote{With respect to ref. \cite{DD} we include the QED $\beta$-function contribution in the definition of $d_e$. Also we are using the same normalization conventions for the $U(1)$ and $SU(3)$ gauge fields.}. 

The QCD frame is defined by performing a Weyl rescaling of the form (\ref{linear})  with $\alpha = -d_g$. This way, the coupling of the dilaton to the gluons disappears from the action.  The general QCD-frame matter Lagrangian, by including dilaton couplings up to linear order, reads
\begin{align} \nonumber
{\cal L}^{(qcd)}_ m = &- \frac{1}{4 g_3^2}\,  F^a_{\mu \nu} F^{a\, \mu \nu} - \frac{1}{4 e^2} \left(1 +  c_e {2 \beta(e)\over e} \varphi\right) F_{\mu \nu} F^{\mu \nu} \\ 
& + \!\! \sum_{i=u,d,e}\left[i{\bar \psi}_i \, {\slash \!\!\!\!D} \, \psi_i - m_i(1+ c_{m_i}\varphi) {\bar \psi}_i \psi_i\right]\, . \label{qcdaction}
\end{align}
where 
\be \label{ci}
c_e = d_e - d_g,\quad c_{m_i} = d_{m_i} - d_g\, .
\ee

Notice that, if all the $d_i's$ were equal, as already noted in \cite{DD} the dilaton would couple to the trace of the stress tensor, and the E-frame action would take the form (\ref{fullaction}). In this case the theory would be  metric and the  QCD frame would reduce to the J-frame.

The gravitational action in the QCD frame reads:
\begin{eqnarray}
S_G = \frac{M^2}{2} \int d^4 x \sqrt{g} \, \left[(1-2 \alpha \varphi) R - 2 \partial_\mu \varphi \partial^\mu \varphi \right]\  , 
 \end{eqnarray}
where we have included the kinetic term of a suitably normalized minimally coupled massless scalar. Since we focus on such a QCD-metric from now on, we indicate it simply as $g_{\mu \nu}$ without the concern of confusing it with the E-metric of eq.~\eqref{emetric}.

Electromagnetic effects and the `bare' masses of quarks and electrons contribute to the total mass of an atom only by a factor of about $10^{-3}$~\cite{DP,DD}. Hence, in first approximation, atoms follow trajectories that are geodesics for the QCD-metric $g_{\mu \nu}$. In this sense, the QCD frame is a good replacement for the J-frame in the case of non-metric theories. More subtle is to understand what `times' and `distances' do our instruments measure in the case of a non-metric theory -- a separate analysis should be done for each measuring device. It is plausible that measurements of lengths and time intervals are generally more sensitive to the electromagnetic and electron-mass couplings $c_e$ and $c_{m_e}$. 

Since the universal (which is also the most sizable) part of the contribution to nucleon masses has been removed, the coefficients $c_i$ in the QCD frame action (\ref{qcdaction}) parametrize directly the purely {\em composition dependent} effects in tests of the equivalence principle: on the other hand these tests cannot constrain general dilaton couplings as they appear in the Einstein frame. 
The introduction of the QCD frame makes manifest the result obtained in \cite{DD}, who showed by direct calculations that only the differences $d_i-d_g$ contribute to composition-dependent violations of the EP. 

In the remaining part of this section we find the deviations from the geodesic motion for test bodies of different compositions in the QCD-frame. By varying the action of a test particle of $\varphi$-dependent mass $m_A$, $S_A = -\int \sqrt{g_{\mu \nu}dx^\mu dx^\nu}m_A(\varphi)$, we find a proper acceleration proportional to the gradient of $m_A(\varphi)$.  If the initial velocity is null (i.e. in the reference frame that is `almost' free falling with the particle), the proper acceleration reads
\begin{equation} \label{acce}
\delta {\vec a}_A = - \frac{d \log m_A}{d \varphi} \, {\vec \nabla} \varphi\, .
\end{equation}
Note that in the QCD-frame only QED and quark and electron masses are responsible for the `effectively varying mass' of an atom. These different contributions are directly proportional to the relative couplings $c$,
\begin{equation} \label{bara}
{\bar \alpha}_A \equiv \frac{d \log m_A}{d \varphi} = Q^{(A)}_e \, c_e + Q^{(A)}_{m_i} \, c_{m_i}\, .
\end{equation}
In the above expression a sum over $i=u,d,e$ is implied in the last term on the RHS. The `charges' of the species $A$, defined as,
\begin{equation}
  Q^{(A)}_e = \frac{\partial \log m_A}{\partial \log e^2}, \qquad Q^{(A)}_{m_i}  = \frac{\partial \log m_A}{\partial \log m_i}\, 
  \end{equation}
are accurately estimated in~\cite{DD} as a function of the atomic number $Z$ and the nucleon number $A$. 

The atomic dilaton couplings ${\bar \alpha}_A$ defined in~\eqref{bara} should be compared with their more often used E-frame counterparts 
\begin{equation}
\alpha_A = \alpha + {\bar \alpha}_A\, .
\end{equation}
The common -- and typically dominant -- term $\alpha$ just gives a constant non-geodesic drift to all particles in the E-frame, independently of their compositions. 
On the opposite, particles follows geodesics `on average' in the QCD-metric.

EP experiments measure the fractional difference of acceleration 
of two test particles $A$ and $B$ falling in the gravitational field of a common attractor, 
\begin{equation}
\left(\frac{\Delta a}{a}\right)_{AB} = \, 2\,  \frac{|\vec a_A - \vec a_B|}{| \vec a_A + \vec a_B|}\, .
\end{equation}
Such accelerations should be intended with respect to a Newtonian reference frame. At the Newtonian order of approximation,
\begin{equation}
 {\vec a}_A = - \vec \nabla V_{\rm New} + \delta {\vec a}_A
 \end{equation}
 where the last term on the RHS is the composition-dependent scalar contribution~\eqref{acce}. The Newtonian potential and the scalar field shape are both generated by a common attractor in the proportion\footnote{See e.g.~\cite{D} -- this is better seen in the E-frame where the propagators of the scalar field and the gravitons decouple} $\varphi = \alpha V_{\rm New}$. We conclude that 
\begin{equation}
\left(\frac{\Delta a}{a}\right)_{AB} = \alpha(\bar \alpha_A - \bar \alpha_B).
\end{equation}
Since this is a frame-invariant quantity, not surprisingly, we recover a known result.

\section{Frame stability and dilaton coupling running} \label{sec:4}

Now we want to address the question whether the `QCD-frame' prescription -- that the dilaton coupling to the gluons be null -- is RG invariant and, more generally, how the dilaton couplings $d_i$ and $c_i$ run with the scale in any given frame. 

We start with the Einstein frame. One may consider the terms in the action~\eqref{efull} 
as $\varphi$-dependent gauge couplings and masses~\cite{DP}. 
Let  us consider for definiteness the QCD coupling $g_3$, which for simplicity we denote $g$ in this section (similar remarks hold for the electromagnetic coupling). From equation \eqref{efull} we can regard $g$  a $\varphi$-dependent coupling:
\begin{equation}\label{27}
g(\varphi) = g_{0} \left(1 - \frac{d_g \beta}{g_0} \varphi\right)\, ,
 \end{equation}
where $g_0$ is $\varphi$-independent and  corresponds to the QCD coupling in a trivial dilaton background. We can  extract the parameter $d_g$ by differentiating w.r.t. $\varphi$: 
\begin{equation} \label{31}
d_g = - \frac{g}{\beta} \frac{\partial \ln g}{\partial \varphi}  .
\end{equation}
The action \eqref{efull} is written in terms of the coupling at the UV cutoff  but we can also define a running $d_g(\mu)$ and write equation (\ref{27}) at any scale $\mu$ by substituting the cutoff coupling for the running coupling. 
As long as no mass thresolds are crossed, it is easy to see that $d_g$ is in fact independent of scale: by  differentiating~\eqref{31} with respect to $\log \mu$ we obtain: 
 \begin{equation} \label{29}
\frac{ \partial \, d_g}{\partial \log \mu} = 0 \, . \qquad \ \ 
\end{equation}  
To obtain this result it is crucial to assume that the only dilaton-dependence of the $\beta$-function is through $g$. We will see below that this is generally invalid when one  crosses a physical mass threshold and integrates out a massive particle. 

The running of mass-dilaton couplings $d_m$ can be studied by analogous means. By the definition, the anomalous dimension $\gamma$ appearing in \eqref{efull} is given by 
\be \label{311}
{d \log m \over d \log \mu} = -\gamma(g), 
\ee
which we can integrate in the form 
\begin{equation}\label{312}
\ln m(\Lambda_1) - \ln m(\Lambda_2) = \int_{g(\Lambda_1)}^{g(\Lambda_2)} \frac{\gamma dg}{\beta}\, .
\end{equation}
The masses at different scales can be considered as functions of $\varphi$: for example at the UV-cutoff we have $m(\Lambda_{UV}) = m( 1 + d_m \varphi + \gamma d_g \varphi)$. 
Differentiating equation (\ref{312})  with respect to $\varphi$ and  assuming that both $\Lambda_1$ and $\Lambda_2$ are dilaton-independent, we obtain: 
\begin{equation} \label{313}
{d \log m(\Lambda_1)  \over d \varphi} -  d_g \gamma(\Lambda_1) = {d \log m(\Lambda_2)  \over d \varphi} -  d_g \gamma(\Lambda_2), 
\end{equation}
where we have made use of equation \eqref{31}. The above equation implies  that, at any scale $\Lambda$, 
\be\label{314}
{d \log m(\Lambda)  \over d \varphi} = d_m + \gamma(\Lambda) d_g, 
\ee
where  the coefficient $d_m$ does not depend on $\varphi$,  and the only scale-dependence of the dilaton couplings in (\ref{efull}) is through that of their anomalous dimensions. In other words, we re-derived the fact that the coefficients $d_g$ and $d_m$ are defined to be RG-invariant~\cite{DD}. 

There is however a subtlety that we encounter when, along the RG-flow, we are in the vicinity of the mass of a particle: in this case, the $\beta$-function will also be sensitive  to the ratio of the RG-scale to that mass. For example,  in crossing a mass threshold, the one-loop $\beta$-function coefficient changes to account for the lower number of degrees of freedom at low energy, and becomes approximately constant again when we are far below the mass of the particle we have integrated out. 

For simplicity, we will consider the RG-running at the one-loop order. The RG equation for the ($\varphi$-dependent) gauge couplings between two different scales $\Lambda_1$ and $\Lambda_2$ can be integrated to give the familiar relation
\begin{equation} \label{flow0}
\frac{1}{g^2(\Lambda_1)} - \frac{1}{g^2(\Lambda_2)} = - b_{N_c,N_f} \ln \frac{\Lambda_1^2}{\Lambda_2^2}, \quad b_{N_c,N_f} = {11 N_c-2 N_f \over 48\pi^2}
\end{equation}

Let us turn to the case in which between $\Lambda_1$ and $\Lambda_2$ there is one charged particle whose mass $m_*$ depends non-trivially on $\varphi$, say $\Lambda_2 \ll m_* \ll \Lambda_1$. The parameter $b$ is now a function of the scale. In particular, it will smoothly vary across $\mu = m_*$ and asymptote to, say, $b_1 =   b_{N_c,N_f}$ in the UV, and $b_2 =   b_{N_c,N_f-1}$ in the IR.
 Accordingly,  the $\beta$ function depends explicitly on $m_*$:
\begin{equation} \label{28}
\beta(\mu) = b\left(\frac{\mu^2}{m^2_*}\right) g^3(\mu)\, .
\end{equation}
Then,~\eqref{flow0} generalizes to 
\begin{equation} \label{flow1}
\frac{1}{g^2(\Lambda_1)} - \frac{1}{g^2(\Lambda_2)} = -  \int_{\Lambda_2^2/m^2_*}^{\Lambda_1^2/m^2_*} \!\! dx \, \frac{b(x)}{x}\, .
\end{equation}
To estimate  the effect of crossing the mass $m_*$ we differente~\eqref{flow1} with respect to $\varphi$ on both sides\footnote{We should specify the dilaton dependences of the two mass scales $\Lambda_1$ and $\Lambda_2$. We will simply assume them to be $\varphi$-independent in the frame of interest, so that the Wilsonian UV and IR scale keep the ``background'' meaning they have in the absence of $\varphi$ }. By making use of~\eqref{31} and \eqref{28} this gives the ``jump'' of the gauge-dilaton coupling $d_g$ when we integrate out one quark of physical mass $m_*$:
\begin{equation} \label{kick}
d_g(\Lambda_2) b_2 = d_g(\Lambda_1) b_1 + {d \log m_* \over d\varphi} (b_2 - b_1)\, .
\end{equation}
To obtain the dependence of $m_*$ from $\varphi$ we have to specify $m_*$. It is natural to define it  as the RG-invariant mass where the RG-scale $\mu$ crosses the quark mass  $m(\mu$), 
\be  
m(m_*) = m_*. 
\ee
Generically, the $\varphi$-dependence of the cutoff mass parameter will give a similar dependence on $\varphi$  in $m_*$. We can obtain this explicitly by rewriting eq. (\ref{312}) and replacing  $\Lambda_1$ with $m_*$ and $\Lambda_2$ with the $\varphi$-independent  UV cutoff $\Lambda_{UV}$.  In this case however the lower integration limit does depend on $\varphi$, so the result (\ref{313}) picks up an extra term: 
\begin{equation} \nonumber
{d \log m(m_*) \over d\varphi} = {d \log m(\Lambda) \over d\varphi} - d_g \, [\gamma(\Lambda) - \gamma_* ]  - \gamma_* {d \log m_* \over d\varphi}, 
\end{equation}
where we have defined $\gamma_* = \gamma(m_*)$. Using the definition of $m_*$ and equation (\ref{314}) we obtain: 
\be \label{40}
{d \log m_* \over d\varphi}
 = d_g + {d_m - d_g \over 1+  \gamma_*} .
\ee 
With this result, we can finally use equation \eqref{kick} to evaluate the ``kick'' in the gauge-dilaton coupling $d_g$ when crossing a mass threshold in terms of the calculable parameters $b_{N_c,N_f}$:
\begin{equation} \label{kick2}
d_g(\Lambda_2) = d_g(\Lambda_1) + \frac{b_2 - b_1}{b_2 (1+\gamma_*)} \left[d_m - d_g(\Lambda_1)\right]\, .
\end{equation}

In metric theories the second term in equation (\ref{kick2}) vanishes and  $d_g$ is unchanged by integrating out a heavy quark:  the 
change in the dilaton-gluon coupling is completely encoded  in the change of the beta function coefficients. As a result, the action will take the form (\ref{fullaction}) at the scale $\Lambda_2$, with the same dilaton coupling coefficients $d_g=d_{m_i}=d_e=-\alpha$, but with the appropriate beta function relevant for the degrees of freedom at that scale. In other words, for a metric theory the dilaton coupling coefficients in the Einstein frame do not depend on the scale. Furthermore, 
if we now perform the Weyl tranformation backwards  at the scale $\Lambda_2$, with the {\em same} coefficient $\alpha$ we used at the scale $\Lambda_1$, we obtain again the action in the Jordan frame, with no explicit dilaton couplings to matter at the scale $\Lambda_2$. This is exactly what we would have expected to find  by running the RG-flow from $\Lambda_1$ to $\Lambda_2$  {\em directly in the Jordan frame,} with a $\varphi$-independent  quark mass scale $m_*$. This example shows explicitly that, as we argued in the introduction, metricity is stable under the RG-flow even when the particles are being integrated out. 

The same considerations apply to the electromagnetic coupling. Let us consider the QED dilaton coupling $d_e$ in the lowest energy limit, i.e. at scales $\mu \ll m_e$. This limit could be of interest for experiments and observations of $\alpha_{QED}$ variations. In this regime $b$ and $\beta$ go to zero linearly in $\mu /m_e^2$. Thus, it is better to define the dilaton coupling $\tilde d_e$ as in~\eqref{defi} i.e. without multiplying it for the combination $\beta/e$. By applying equation~\eqref{kick2} we then get 
\begin{equation} \label{kick3}
\tilde d_e(\mu) =  \frac{2 \beta(\mu) }{e} d_{m_e} + 2 e^2 b(\Lambda)  [d_e(\Lambda) - d_{m_e}] \, ,
\end{equation}
where $\Lambda \gg m_e$. If we consider a metric theory at energies $\mu \ll m_e$ and go to the E-frame we still obtain ~\eqref{fullaction}, but now the QED dilaton coupling is suppressed by the vanishing of the $\beta$ function as $\mu^2/m_e^2$. The very same operation could be done by switching to the E-frame at energies above $m_e$ and RG-running down in the E-frame. At those energies, since the theory is metric, $d_e =  d_{m_e}$, and by~\eqref{kick3} we get to the same low energy dilaton coupling. The operation of frame-switching commutes with the RG-running. 

In non-metric theories  things are different: if at the UV scale  $\Lambda_1$ the couplings to quarks are not tuned to those of gluons from the start, as in the general action (\ref{efull}), integrating out some of these  quarks will induce a shift in the dilaton gluon coupling $d_g$. However, according to equation (\ref{kick2}) this change is proportional to the {\em deviation from metricity}, which is controlled at the scale $\Lambda_1$ by the difference $d_m - d_g \equiv c_m$, i.e. by the coupling appearing naturally in the QCD-frame.

 Also, from equation (\ref{314}) it is easy to see that the change in any $c_{m_i}$ when as we run down with the RG flow is at most proportional to other $c_m$'s, since 
$c_{m_i}(\Lambda_2) - c_{m_i}(\Lambda_1 ) = d_g(\Lambda_1) - d_g(\Lambda_2)$.  
Thus, if all the $c_m$'s are small, their change will be small as we move along the RG-flow. Since the coefficients  $c_m$ parametrize the deviation from metricity, we can define a theory to be {\em approximately metric} if all $c_{m_i} \ll 1$. Then the discussion above   shows that {\em approximate metricity   is  stable under (matter) quantum corrections,} which a priori is far from trivial. One can assert this by stating that approximate metricity is  natural in the technical sense. 
In fact,  the matter action in metric theories has an extra exact symmetry, which  protects  metricity against quantum corrections: its infinitesimal action is: 
\be
g_{\mu\nu} \to (1 + 2\epsilon) g_{\mu\nu}, \quad  \psi_i \to (1- 3\epsilon/2)\psi_i, \quad  \varphi \to \varphi - \epsilon/\alpha,   
\ee
where  $\alpha$ is the common dilaton coupling.
 The same (approximate) symmetry is  what protects approximately metric theories
from growing  large deviations from metricity by RG-flow. 
This symmetry however is broken by the Einstein-Hilbert action, thus violations of metricity will be present due to Planckian effects.  

Finally,   let us consider a general non-metric theory in our QCD-frame, in which we have chosen to set $c_g = 0$ at the scale  $\Lambda_{QCD}$,  as apparent from~\eqref{qcdaction}. In this case, according to  equation (\ref{kick2}), as we move up with the energy scale, heavier particles will contribute to the beta function and their dilaton couplings $c_m$ will generate a gluon-dilaton coupling $c_g\neq 0$ according to~\eqref{kick2}. In principle, at a different scale, another conformal transformation is needed to re-define a QCD frame, so the definition of the QCD frame is scale-dependent. However in practice, the gravitational phenomenology is most  sensitive to the scale $\mu \sim \Lambda_{QCD}$, so it is at this scale that it is most convenient to set to zero the gluon-dilaton coupling.

\section{Discussion}
In this note we clarified some issues related with scalar-tensor theories when quantum corrections are taken into account, and proposed a conformal frame, the `QCD-frame', that is particularly appropriate to study the violation of the weak equivalence principle in the motion of test  bodies. Our analysis confirms the unavoidable presence of scalar couplings to gauge fields in the E-frame for metric theories different than GR~\cite{brax,cliff}. Following the arguments raised 
in~\cite{brax}, one could then wonder whether the photon-scalar coupling that is generated has observable consequences---weak EP-violations, $\alpha_{QED}$ variations etc\dots \,. We took the ``privileged'' J-frame viewpoint to argue that none of those effects are in order. More generally, E-frame dilaton couplings have no universal implication on EP-phenomenology.

A point which is worth stressing is that, although all frames are physically equivalent, different frames can be more convenient to address different questions. The fact that in the Einstein and Jordan (or QCD) frame the dilaton couplings  in the Lagrangian are different, does not mean that the answer to physical questions is different. For example consider, say, the production of photons in a time-dependent dilaton background, but in which the Einstein frame metric is static. In the Einstein frame this process will be governed solely by the dilaton-photon coupling, $d_e \beta(e)/e$ of equation (\ref{efull}). Suppose now we go to a Jordan or Jordan-like  frame where this coupling is absent: here, there will be instead a time-dependent conformally flat metric, which will couple to the photon stress-tensor. This will result in a photon production rate which will arise exclusively from the one-loop breaking of conformal invariance,  and will be proportional to the $\beta$-function, giving exactly the same result. Thus the two frames are physically equivalent, and a one-loop effect in one frame is translated in a tree-level effect in another frame. 
  
However, this does not mean that all frames are equivalently suitable for calculating different effects:  for example, the Einstein frame is most useful when one wants to separate the dynamics of the dilaton from that of the metric, at least for what concerns perturbations around flat space, since in the E-frame the kinetic terms fluctuations are diagonal. Thus,  the meaning of E-frame dilaton couplings is most transparent when one wants to deal with quantum processes involving the dilaton: in order to calculate the decay rate of a dilaton, say, into two photons, or the dilaton emission by a photon,  we need to suitably define asymptotic states; and the appropriate \emph{``in"} state for this problem is the E-frame dilaton. In any other frame, due to kinetic mixing, the \emph{``in"} state would be a mixture of dilaton and gravitons. It follows that the E-frame dilaton couplings are those to be used for calculating the decaying rate of such state. 

Similarly, when one is interested solely in the violation of the equivalence principle, we argue that the QCD frame is the most transparent, since it is only in this frame that all couplings are  proportional to weak EP violations,  and therefore have a direct physical meaning as far as   these effects are concerned. 
Possible variations of the coupling constants like $\alpha_{\rm QED}$, or experimental bounds thereof~\cite{uzan,chiba}, should be analyzed case by case depending on the specific instruments adopted. However, we argue that the QCD-frame is, again, a preferred view point to quantify them. Since in the limit of a metric theory $c_e$ -- and not necessarily $d_e$ -- vanishes, those effects are naturally proportional to the parameter $c_e$: the QCD-frame coupling. 

Finally, the QCD couplings, rather than those in the Einstein frame,  are the ones that directly   parametrize small deviations from metricity.  This is particularly important in light of the Renormalization Group analysis we performed in Section IV, where we analyzed the stability of metricity and approximate metricity under radiative corrections and RG-flow. We found that not only exact, but even  {\em approximate} metricity is stable under the RG-flow, 
The definition of approximate metricity involves again the dilaton-matter couplings in the QCD frame: if they are small, then one can call the theory approximately metric. Here, we showed that this property is technically natural: if valid at the cutoff, it is preserved under RG-flow and radiative corrections. 
For metric theories, this statement reduces to the expected result that metricity is preserved under RG-flow, as could be argued
on general grounds  by performing calculations directly in the Jordan frame.

 \vspace{-0.5cm}
\acknowledgments We are grateful to Philippe Brax, Cliff Burgess, John Donoghue, Elias Kiritsis, Jihad Mourad, Alberto Nicolis, Lam Hui and David Seery for discussions and correspondence.

\appendix
\section{The `physicality' of Jordan frame}\label{app_a}

In this appendix we clarify the statement made in the introduction that the Jordan frame is the ``physical frame". In field theory we are used to field redefinitions that leave the physical content of a theory (i.e. scattering amplitudes) unchanged. However, in GR the metric tensor $g_{\mu \nu}$ has also the straightforward semi-classical meaning of physical distance, in the sense that, integrated along a curve, it gives the physical length of such a curve. To be more precise, a curve parametrically defined by $x^\mu =  x^\mu(\lambda)$, with $\lambda$ a real parameter evaluating $\lambda_1$ and $\lambda_2$ at the two extremes, has length 
\begin{equation} \label{distance}
l = \int ds = \int_{\lambda_1}^{\lambda_2} \sqrt{g_{\mu \nu} \frac{d x^\mu}{d \lambda} \frac{d x^\nu}{d \lambda} } d\lambda\, .
\end{equation}
This can be a space distance or a time interval. Now, it is an interesting question whether or not $l$ is measurable in practice. If the curve is timelike and smooth enough it simply can be the world-line of a clock. Space-like curves are more subtle,  but certainly $l$ has a very well defined operational meaning in a stationary spacetime. Stationarity allows to make a local operation (i.e. lay down a ruler) several times in the same physical conditions. The measurement, in practice, takes place along a timelike or null trajectory, but this is fine because of stationarity: we can project the events $x^\mu(\lambda_1)$ and $x^\mu(\lambda_2)$ arbitrarily far in the past and future along the time-like Killing vector field.  

For scalar tensor theories the problem poses as which one of the different conformally related metrics, when used in~\eqref{distance}, gives the measured physical length $l$. If there exists a metric that minimally couples to matter fields it is easy to convince oneself that that is the metric that we are talking about. 
Imagine, for definiteness, that there is a strong gradient of scalar field between $x^\mu(\lambda_1)$ and $x^\mu(\lambda_2)$, so that the Einstein and Jordan metrics, when used in eq.~\eqref{distance}, produce two very different outcomes $l^{(E)}$ and $l^{(J)}$. The matter field dynamics, responsible for the internal equilibrium of the atoms inside the rod, is insensitive to such a gradient only in the Jordan frame. This means that only $l^{(J)}$ will be proportional to the number of rods fitting between $x^\mu(\lambda_1)$ and $x^\mu(\lambda_2)$, which is the chosen operational definition of length. Similar considerations can be drawn for measures of time intervals. 
Of course, by previous and independent knowledge of the scalar field's configuration we can easily switch from $l^{(J)}$ to $l^{(E)}$ and in this respect the two frames are perfectly equivalent. But the statement that, say, the mean earth-moon distance is 384,400 km implies a measurement procedure that does not take into account any scalar field and therefore is a statement about the Jordan length $l^{(J)}$.

Cosmology offers another useful example for this discussion.  The metric of a spatially flat Friedman Robertson Walker reads, say, in Einstein frame, $ds^2 = - dt^2 + a^2(t) dx^2$. Upon conformally transform to Jordan frame, $\tilde{g}_{\mu\nu} = \Omega^2(\phi) g_{\mu\nu}$, and an appropriate redefinition of the time coordinate, the metric becomes 
\begin{equation}
ds_J^2  = - dt_J^2 + \Omega^2a^2(t)  dx^2
\end{equation}
which defines a different scale factor, $a_J(t) = \Omega (\phi(t))a(t)$ and therefore a different Hubble constant $H_J = d \ln a_J/ d t_J$.
Because in Jordan frame Maxwell's equations are oblivious to $\phi$, the photons from a far away source simply redshift with the Jordan frame scale factor. This means that in a scalar tensor theory the redshift has the usual expression in terms of the scale factor, only if the J-frame is used for the latter, 
\begin{equation}
1 + z = \frac{a_J(t)}{a_J(t_0)}\, , 
\end{equation}
$t_0$ being the present time. 
With a similar straightforward argument, just more pedantic, we could also conclude that the value of the Hubble constant measured by the Hubble Space Telescope~\cite{h0}, $H_0 = 72 \, \pm \, 8 \, {\rm km} / {\rm s}\, {\rm Mpc}^{-1}$, matches that of the Jordan frame quantity $H_J$. 
Again, we can equivalently use $H_E = d \ln a/dt$ instead, but in order to determine the value of the latter we need also independent knowledge of $\phi$ and $\dot \phi$.



\vspace{-0.5cm}


\end{document}